\documentclass[10pt,final,doublecolumn]{IEEEtran}
\usepackage{amsmath}
\usepackage{latexsym}
\usepackage{graphicx}
\usepackage{bbding}
\usepackage{indentfirst}
\usepackage{algorithm,algorithmic}
\usepackage{setspace}
\usepackage{float}
\usepackage{epstopdf}
\usepackage{amssymb}
\usepackage{amsfonts}
\usepackage{enumerate}
\usepackage{multicol}
\usepackage{color}
\usepackage{slashbox}
\usepackage{bm}
\usepackage{amssymb}
\usepackage{stfloats}
\usepackage{epstopdf}
\usepackage{threeparttable}
\usepackage{epstopdf}
\usepackage{threeparttable}
\usepackage{subfigure}

\usepackage{cite}

\IEEEoverridecommandlockouts
\allowdisplaybreaks[4]

\begin{document}
\title{Exploiting Six-Dimensional Movable Antenna (6DMA) for  Wireless Sensing}


%
%
%
%

\author{{Xiaodan Shao,~\IEEEmembership{Member,~IEEE}, Rui Zhang, \IEEEmembership{Fellow, IEEE}, and Robert
Schober,~\IEEEmembership{Fellow, IEEE}}

\thanks{X. Shao and R. Schober are with the Institute for Digital Communications, Friedrich-Alexander-University Erlangen-Nurnberg (FAU), 91054
Erlangen, Germany (e-mails:xiaodan.shao@fau.de, robert.schober@fau.de).}	
\thanks{R. Zhang is with School of Science and Engineering, Shenzhen Research Institute of Big Data, The Chinese University of Hong Kong, Shenzhen, Guangdong 518172, China. He is also with the Department of Electrical and Computer Engineering, National University of Singapore, Singapore 117583 (e-mail: elezhang@nus.edu.sg).}
}
\maketitle
\IEEEpeerreviewmaketitle

\begin{abstract}
Six-dimensional movable antenna (6DMA) is an emerging technology
that is able to fully exploit the spatial variation of wireless channels by controlling the 3D positions and 3D rotations of distributed antennas/antenna surfaces at the transmitter/receiver.
In this letter, we apply 6DMA at the base station (BS) to enhance its wireless sensing performance over a given set of regions. To this end, we first
divide each region into a number of equal-size subregions
and select one typical target location within each subregion. Then, we derive an expression for the Cramer-Rao bound (CRB) for estimating the directions of arrival (DoAs) from these typical target locations in all regions, which sheds light on the sensing performance of 6DMA enhanced systems in terms of a power gain and a geometric gain.
Next, we minimize the CRB for DoA estimation via jointly optimizing the positions and rotations of all 6DMAs at the BS, subject to practical movement constraints, and propose an efficient algorithm to solve the resulting non-convex optimization problem sub-optimally. Finally, simulation results demonstrate the significant improvement in DoA estimation accuracy achieved by the proposed 6DMA sensing scheme as compared to various benchmark schemes, for both isotropic and directive antenna radiation patterns.
\end{abstract}

\begin{IEEEkeywords}
6DMA, antenna position and rotation optimization, wireless sensing, DoA estimation, Cramer-Rao bound (CRB).
\end{IEEEkeywords}

\section{Introduction}
Next-generation wireless networks will support many
new applications such as autonomous transportation, ubiquitous sensing, extended reality
and so on, which demands not only higher-quality wireless
connectivity but also unprecedentedly higher sensing accuracy
than what is provided by today's wireless systems. Motivated by this, integrated sensing and communication (ISAC) has been
recognized as an essential technology for the sixth-generation (6G) wireless network \cite{liufan, shaos}. However, in practical scenarios with complex radio propagation environments, the performance of sensing may degrade significantly, which limits the fundamental trade-off between sensing and communication performance in ISAC systems.
\begin{figure}[t!]
\vspace{-0.69cm}
\centering
\setlength{\abovecaptionskip}{0.cm}
\includegraphics[width=2.6in]{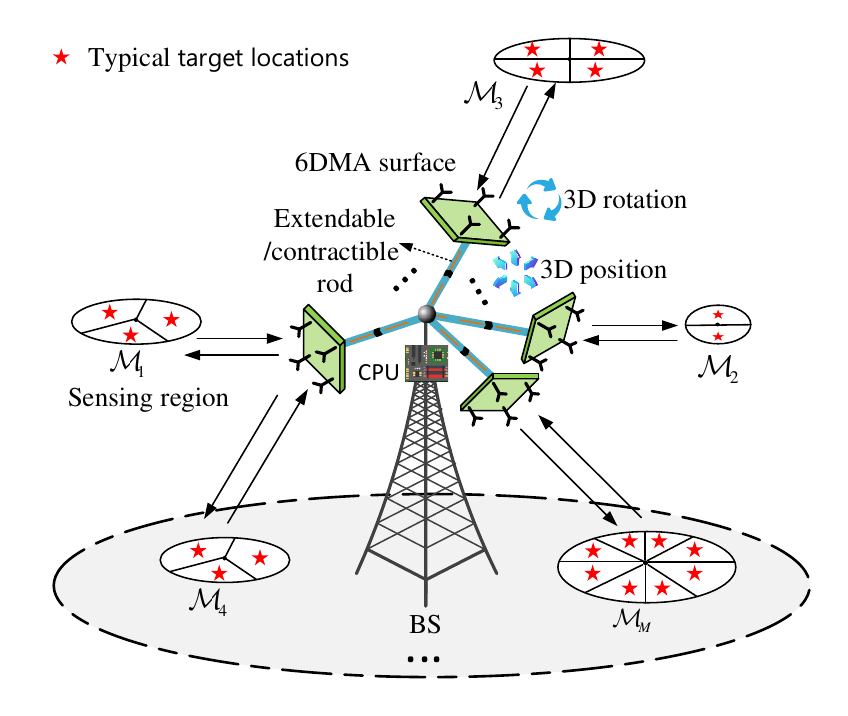}
\caption{6DMA-enabled BS adaptable to different sensing regions.}
\label{practical_scenario}
\vspace{-0.69cm}
\end{figure}

Recently, to fully exploit the spatial variation of wireless channels
at the transmitter/receiver, six-dimensional movable antennas (6DMAs) comprised of multiple rotatable and positionable antennas/antenna surfaces has been
proposed as a new technology to improve the performance of future wireless networks without increasing the number of antennas \cite{shao20246d, 6dma_dis,shaomaga,jstsp}. Equipped with distributed 6DMAs to match the spatial user distribution, 6DMA-enabled transmitters/receivers can adaptively allocate antenna resources in space to maximize the array gain, and spatial multiplexing gain while also suppressing interference. This thus significantly enhances the capacity of wireless communication networks.
In practice, the positions and rotations of 6DMAs can be adjusted continuously \cite{shao20246d} or discretely \cite{6dma_dis} based on the surface movement mechanism, and they can be optimized with or without prior knowledge of the spatial channel distribution of the users in the network \cite{shao20246d, 6dma_dis}.

Although the benefits of 6DMAs in wireless communication
systems have been validated, their potential performance gains
in wireless sensing systems remain to be explored, which
motivates this letter. As shown in Fig. \ref{practical_scenario},
we consider a 6DMA-enabled wireless sensing system where the base station (BS) is equipped with distributed antenna surfaces whose 3D positions and 3D rotations can be flexibly adjusted to enhance the sensing performance over a set of distributed regions. We partition each region into a number of equal-sized subregions, then select one typical target location within each subregion.
For the above setup, we first demonstrate that the performance enhancement achieved by the proposed 6DMA sensing system comprises both a power gain and a geometric gain. Next, we minimize
the Cramer-Rao bound (CRB) for estimating the directions of arrival (DoAs) of the typical target locations by jointly optimizing the 6DMAs' positions and rotations at the BS subject to a set of practical movement constraints. We propose an efficient algorithm
to solve the formulated non-convex optimization problem sub-optimally.
Numerical results reveal the significant performance improvement achieved by our proposed 6DMA sensing scheme over a fixed-position antenna (FPA) system and fluid antenna (FA)/ movable antenna (MA) systems \cite{9650760,zhulet,weidong}, for both isotropic and directive antenna radiation patterns. Note that FA/MA adjusts the positions of active antennas on a finite 2D surface only to exploit small-scale
channel fading, without considering the antennas' rotation.

\emph{Notations}: $(\cdot)^*$, $(\cdot)^H$, and $(\cdot)^T$  respectively denote conjugate, conjugate transpose, and transpose, $\mathbb{E}[\cdot]$ denotes expectation of a
random variable,  $\left \|\cdot\right \|_2$ denotes the Euclidean norm, $\mathrm{diag}({\bf x})$ denotes a diagonal matrix with the diagonal entries specified by vector ${\bf x}$, $[\mathbf{x}]_{j}$ denotes the $j$-th element of vector $\mathbf{x}$, $[\mathbf{A}]_{:,j}$ denotes the $j$-th column of matrix $\mathbf{A}$, $\mathrm{Re}\{\cdot\}$ denotes the real part of a complex number, $\mathrm{tr}(\mathbf{A})$
denotes the trace of square matrix $\mathbf{A}$, and $|\mathcal{X}|_{\mathrm{c}}$ denotes the cardinality of set $\mathcal{X}$.

\section{System Model}
As shown in Fig. \ref{practical_scenario}, we consider a wireless sensing network with a 6DMA-BS for sensing $M$ distributed regions, denoted by $\mathcal{M}_m, m=1,2,\cdots,M$. We assume that all regions have a common sensing accuracy requirement.
We divide region $\mathcal{M}_m$ into $K_m$ subregions of equal size $\Delta$ such that the $K_m, m=1,\cdots,M$, are proportional to the areas of their corresponding regions. Next, we select the central position within each subregion as the typical target location for it. We assume that the wireless channels from the BS to all typical target locations are line of sight (LoS).

\subsection{6DMA Architecture}
The 6DMA-BS is assumed to employ a mono-static radar (i.e., radar transmitter and receiver are co-located), which is equipped with $B$ 6DMA surfaces, whose indices are collected in set $\mathcal{B} = \{1, 2, \ldots, B\}$. Each 6DMA surface is assumed to be a  uniform linear array
(ULA) with a given size, which comprises $N\geq 1$ antennas, indexed by  set $\mathcal{N} \triangleq \{1, 2, \ldots, N\}$\footnote{We consider a ULA for simplicity as, in this letter, we focus on the estimation of the DoAs of targets in the horizontal direction. Our results can be extended to the more general setup with 3D DoA estimation for both aerial and terrestrial targets (see Fig. 1).}. The 6DMA surfaces are connected to a central processing unit (CPU) at the BS via rods that can be extended or retracted and are equipped with flexible wires, and thus their 3D positions, denoted by $\mathbf{q}_b=[x_b,y_b,z_b]^T\in\mathcal{C}$, and 3D rotations, denoted by $\mathbf{u}_b=[\alpha_b,\beta_b,\gamma_b]^T$, can be adjusted by the CPU.
Here, $\mathcal{C}$ denotes the given 3D space at the BS for the movement of the 6DMA surfaces; $x_b$, $y_b$, and $z_b$ denote the coordinates of the center of the $b$-th 6DMA surface in the global Cartesian coordinate system (CCS) $o\text{-}xyz$, where $o$ is the origin located at the BS's reference position. The angles $\alpha_b$, $\beta_b$, and $\gamma_b$ lie in the intervals $[0, 2\pi)$ and represent rotations around the $x$-, $y$-, and $z$-axes, respectively.

Given $\mathbf{u}_b$, the following rotation matrix can be defined:
\begin{align}\label{R}
&\!\!\!\mathbf{R}(\mathbf{u}_b)\nonumber\\
&\!=\!\begin{bmatrix}
c_{\alpha_b}c_{\gamma_b} & c_{\alpha_b}s_{\gamma_b} & -s_{\alpha_b} \\
s_{\beta_b}s_{\alpha_b}c_{\gamma_b}-c_{\beta_b}s_{\gamma_b} & s_{\beta_b}s_{\alpha_b}s_{\gamma_b}+c_{\beta_b}c_{\gamma_b} & c_{\alpha_b}s_{\beta_b} \\
c_{\beta_b}s_{\alpha_b}c_{\gamma_b}+s_{\beta_b}s_{\gamma_b} & c_{\beta_b}s_{\alpha_b}s_{\gamma_b}-s_{\beta_b}c_{\gamma_b} &c_{\alpha_b}c_{\beta_b} \\
\end{bmatrix},\!\!
\end{align}
where $c_{x}=\cos(x)$ and $s_{x}=\sin(x)$  \cite{shao20246d,6dma_dis}. Let $\bar{\mathbf{r}}_{n}$ denote the position of the $n$-th antenna of any 6DMA surface in its local CCS. Then, the position of the $n$-th antenna of the $b$-th 6DMA surface in the global CCS can be described as
\begin{align}\label{nwq}
\mathbf{r}_{b,n}(\mathbf{q}_b,\mathbf{u}_b)=\mathbf{q}_b+\mathbf{R}
(\mathbf{u}_b)\bar{\mathbf{r}}_{n},~n\in\mathcal{N},~b \in\mathcal{B}.
\end{align}

As explained in \cite{shao20246d,6dma_dis}, the following three practical constraints for rotating/positioning 6DMA surfaces need to be considered:
\begin{align}
&\mathbf{n}(\mathbf{u}_b)^T(\mathbf{q}_{j}-\mathbf{q}_b)\leq  0,~\forall b ,j \in \mathcal{B}, j\neq b, \label{rcc}\\
&\mathbf{n}(\mathbf{u}_b)^T\mathbf{q}_b\geq 0,~\forall b\in \mathcal{B}, \label{dd}\\
&\|\mathbf{q}_b-
\mathbf{q}_{j}\|_2\geq d_{\min},~\forall b ,j \in \mathcal{B}, j\neq b, \label{dss}
\end{align}
where $\mathbf{n}(\mathbf{u}_b)=\mathbf{R}(\mathbf{u}_b)\bar{\mathbf{n}}$ denotes the normal vector of the $b$-th 6DMA surface in the global CCS with $\bar{\mathbf{n}}$ denoting the corresponding normal vector in the local CCS.
The minimum distance $d_{\min}$ in constraint \eqref{dss} avoids overlapping and coupling among 6DMA surfaces. Constraint \eqref{rcc} avoids mutual signal reflections between any two 6DMA surfaces, while constraint \eqref{dd} prevents signal blockage by the CPU of the BS.

\subsection{Signal Model}
Let $\phi_k\in[-\pi,\pi]$ denote the horizontal DoA from the $k$-th typical target location to the BS's reference position in the global CCS, where $k\in\{1,2,\cdots,K\}$ with $K=\sum_{m=1}^{M}K_m$.
The pointing vector corresponding to $\phi_k$ is thus given by
\begin{align}\label{KM}
\mathbf{f}(\phi_k)=[\cos(\phi_k), \sin(\phi_k), 0]^T.
\end{align}
By combining \eqref{nwq} and \eqref{KM}, the channel vector for angle $\phi_k$ is modeled as
\begin{align}\label{res}
&\mathbf{h}_k(\mathbf{q},\mathbf{u},{\phi_k})=\left[\!\sqrt{g(\mathbf{u}_1,{\phi_k})}
\mathbf{a}(\mathbf{q}_1,\mathbf{u}_1,{\phi_k})^T,
\!\cdots\!,\right.\nonumber\\
&~~~~~~~~~~~~~\left.\sqrt{g(\mathbf{u}_B,{\phi_k})}\mathbf{a}
(\mathbf{q}_B,\mathbf{u}_B,{\phi_k})^T
\!\right]^T\in \mathbb{C}^{NB\times 1},
\end{align}
where ${\mathbf{q}}=[\mathbf{q}_1^T,\mathbf{q}_2^T,\cdots,\mathbf{q}_B
^T]^T\in \mathbb{R}^{3B\times 1}$, $\mathbf{u}=[\mathbf{u}_1^T,\mathbf{u}_2^T,\cdots,\mathbf{u}_B^T]^T\in \mathbb{R}^{3B\times 1}$, and $\mathbf{a}(\mathbf{q}_b,\mathbf{u}_b,\phi_k)$ denotes the steering vector of the $b$-th 6DMA surface, which is given by
\begin{align}\label{gen}
&\mathbf{a}(\mathbf{q}_b,\mathbf{u}_b,\phi_k)=\nonumber\\
& \left[\!e^{-j\frac{2\pi}{\lambda}
\mathbf{f}(\phi_k)^T\mathbf{r}_{b,1}(\mathbf{q}_b,\mathbf{u}_b)},
\!\cdots,\! e^{-j\frac{2\pi}{\lambda}\mathbf{f}(\phi_k)^T
\mathbf{r}_{b,N}(\mathbf{q}_b,\mathbf{u}_b)}\!\right]^T,
\end{align}
where $\lambda$ denotes the carrier wavelength. In \eqref{res}, $g(\mathbf{u}_b,\phi_k)$ is the effective antenna gain for the $b$-th 6DMA surface in the linear scale, and it is defined as
\begin{align}\label{gm}
g(\mathbf{u}_b,\phi_k)=
10^{\frac{A(\tilde{\theta}_{b},\tilde{\phi}_{b})}{10}},
\end{align}
where
$A(\tilde{\theta}_{b},\tilde{\phi}_{b})$ denotes the corresponding gain in dBi (to be specified in Section V) with $
\tilde{\theta}_{b}=\pi/2-\arccos(\tilde{z}_{b})$, $\tilde{\phi}_{b}=
\arccos(\frac{\tilde{x}_{b}}{\sqrt{\tilde{x}_{b}^2+\tilde{y}_{b}^2}}
)\times\eta(\tilde{y}_{b}) \label{cc}$, $
[\tilde{x}_{b},\tilde{y}_{b},\tilde{z}_{b}]^T=\mathbf{R}(\mathbf{u}_b)^{-1}
\mathbf{f}(\phi_k)=\mathbf{R}(\mathbf{u}_b)^T\mathbf{f}(\phi_k)$, and function $\eta(\tilde{y}_{b})$ returns $1$ if $\tilde{y}_{b} \geq 0$ and $-1$ if $\tilde{y}_{b} < 0$.

The sensing/radar signals transmitted by the 6DMA-BS are collected in matrix $\mathbf{X}\in\mathbb{C}^{NB\times L}$ with
$L>NB$ being the length of the sensing
frame.
By transmitting $\mathbf{X}$ to sense all the typical targets, the received echo signal
matrix at the 6DMA-BS is given by
\begin{align}\label{ly}
\mathbf{Y}&=\mathbf{H}(\mathbf{q},\mathbf{u},\boldsymbol{\phi})\mathbf{Z}
\mathbf{H}(\mathbf{q},\mathbf{u},\boldsymbol{\phi})^H
\mathbf{X}+\mathbf{N},
\end{align}
where $\mathbf{H}(\mathbf{q},\mathbf{u},\boldsymbol{\phi})=[\mathbf{h}_1
(\mathbf{q},\mathbf{u},\phi_1)
, \cdots, \mathbf{h}_K(\mathbf{q},\mathbf{u},\phi_K)]\in \mathbb{C}^{NB\times K}$ with $\boldsymbol{\phi}=[\phi_1,\phi_2,\cdots,\phi_K]^T$ and $\mathbf{Z}=
\text{diag}(\boldsymbol{\rho})$ with $\boldsymbol{\rho}=[\rho_1,\rho_2,\cdots,\rho_K]^T$.
Here, $\rho_k$ represents the complex-valued channel coefficient, which is dependent on the $k$-th target's radar cross section (RCS) and the path loss of the BS-target-BS link. $\mathbf{N}$ denotes an additive white Gaussian noise (AWGN) matrix with each entry having variance $\sigma^2$.

\section{Performance Metric and Problem Formulation}
\subsection{Performance Metric}
In this letter, the 6DMA-BS aims to enhance the performance of target DoA estimation. Thus, for ease of exposition, we assume that
the channel coefficients in $\boldsymbol{\rho}$ have been perfectly estimated, so that we can focus on the problem
of estimating the DoA parameters $\boldsymbol{\phi}$.
Following \cite{crbc}, we can obtain the Fisher information matrix (FIM) for estimating $\boldsymbol{\phi}$, denoted as $\mathbf{F}$, whose element in the $i$-th row and $j$-th column is given by
\begin{align}\label{ff}
\!\!\![\mathbf{F}]_{ij}=\frac{2L}{\sigma^2}\mathrm{Re}\left\{\rho_i^*\rho_j\mathrm{tr}
\left(\dot{\mathbf{H}}
(\mathbf{q},\mathbf{u},\boldsymbol{\phi})\mathbf{S}_{\mathrm{X}}\dot{\mathbf{H}}
(\mathbf{q},\mathbf{u},\boldsymbol{\phi})^H\right)\right\},
\end{align}
where $\mathbf{S}_{\mathrm{X}}=\frac{1}{L}\mathbf{X}\mathbf{X}^H$ and  $\dot{\mathbf{H}}(\mathbf{q},\mathbf{u},\boldsymbol{\phi})=
[\dot{\mathbf{h}}_1(\mathbf{q},\mathbf{u},\phi_1)
, \cdots, \dot{\mathbf{h}}_K(\mathbf{q},\mathbf{u},\phi_K)]$ with $\dot{\mathbf{h}}_k(\mathbf{q},\mathbf{u},\phi_k)=\frac{\partial\mathbf{h}_k(\mathbf{q},\mathbf{u},\phi_k)}{\partial \phi_k}$.

The CRB relevant for estimation of $\boldsymbol{\phi}$ is given by the trace of $\mathbf{F}^{-1}$
\begin{align}\label{CRB}
\mathrm{CRB}\left(\mathbf{q},\mathbf{u},  \boldsymbol{\phi}\right)=\mathrm{tr}(\mathbf{F}^{-1}).
\end{align}

In our considered scenario with multiple targets, the CRB for the estimation of each target's DoA is equal to the corresponding main diagonal element of the inverse FIM, i.e.,
\begin{align}\label{CRBb}
\mathrm{CRB}_k\left(\mathbf{q},\mathbf{u},  \phi_k\!\right)=[\mathbf{F}^{-1}]_{kk}.
\end{align}
To get some insight for algorithm design, we make the idealizing assumption that the DoA estimations of different targets are independent \cite{inner}. Under this assumption, we can derive an explicit expression for \eqref{CRBb}, which is shown in \eqref{Crr} \cite{inner} at the top of the next page,
\begin{figure*}
\begin{align}\label{Crr}
&\mathrm{CRB}_{k}\left(\mathbf{q},\mathbf{u},  {\phi_k}\!\right)\nonumber\\
&=\frac{\sigma^2\text{tr}(\mathbf{A}
(\mathbf{q},\mathbf{u},\phi_k)^H\mathbf{A}
(\mathbf{q},\mathbf{u},\phi_k)\mathbf{S}_{\mathrm{X}})}{2
|\rho_k|^2L(\text{tr}(\dot{\mathbf{A}}
(\mathbf{q},\mathbf{u},\phi_k)^H\dot{\mathbf{A}}
(\mathbf{q},\mathbf{u},\phi_k)\mathbf{S}_{\mathrm{X}})
\text{tr}(\mathbf{A}
(\mathbf{q},\mathbf{u},\phi_k)^H\mathbf{A}
(\mathbf{q},\mathbf{u},\phi_k)\mathbf{S}_{\mathrm{X}})-
|\text{tr}(\dot{\mathbf{A}}
(\mathbf{q},\mathbf{u},\phi_k)^H\mathbf{A}
(\mathbf{q},\mathbf{u},\phi_k)\mathbf{S}_{\mathrm{X}})|^2)}.
\end{align}
\end{figure*}
where $\mathbf{A}(\mathbf{q},\mathbf{u},\phi_k)=
\mathbf{h}(\mathbf{q},\mathbf{u},\phi_k)\mathbf{h}
(\mathbf{q},\mathbf{u},\phi_k)^H$. By calculating the derivative of ${\mathbf{h}}_k(\mathbf{q}, \mathbf{u}, \phi_k)$, it can be verified that the equality ${\mathbf{h}}_k(\mathbf{q}, \mathbf{u}, \phi_k)^H \dot{\mathbf{h}}_k(\mathbf{q}, \mathbf{u}, \phi_k) = 0, \forall \phi_k$ is satisfied if $B=1$ and the center of the 6DMA surface is chosen as the reference point, i.e., the region $o$ \cite{sonar}. Consequently, we have
\begin{align}
&\text{tr}({\mathbf{A}}
(\mathbf{q},\mathbf{u},\phi_k)^H{\mathbf{A}}
(\mathbf{q},\mathbf{u},\phi_k)\mathbf{S}_{\mathrm{X}})\nonumber\\
&=
\|{\mathbf{h}}_k(\mathbf{q},\mathbf{u},\phi_k)\|_2^2\|
{\mathbf{h}}_k(\mathbf{q},\mathbf{u},\phi_k)^H
\mathbf{X}\|_2^2,\label{ju}\\
&\text{tr}(\dot{\mathbf{A}}
(\mathbf{q},\mathbf{u},\phi_k)^H{\mathbf{A}}
(\mathbf{q},\mathbf{u},\phi_k)\mathbf{S}_{\mathrm{X}})\nonumber\\
&=\|{\mathbf{h}}_k(\mathbf{q},\mathbf{u},\phi_k)\|_2^2
{\mathbf{h}}_k(\mathbf{q},\mathbf{u},\phi_k)^H
\mathbf{X}\mathbf{X}^H\dot{\mathbf{h}}_k(\mathbf{q},\mathbf{u},
\phi_k),\label{ju1}\\
&\text{tr}(\dot{\mathbf{A}}
(\mathbf{q},\mathbf{u},\phi_k)^H\dot{\mathbf{A}}
(\mathbf{q},\mathbf{u},\phi_k)\mathbf{S}_{\mathrm{X}})\nonumber\\
&=
\|\dot{\mathbf{h}}_k(\mathbf{q},\mathbf{u},\phi_k)\|_2^2\|
{\mathbf{h}}_k(\mathbf{q},\mathbf{u},\phi_k)^H
\mathbf{X}\|_2^2\nonumber\\
&+\|{\mathbf{h}}_k(\mathbf{q},\mathbf{u},\phi_k)\|_2^2\|
\dot{\mathbf{h}}_k(\mathbf{q},\mathbf{u},\phi_k)^H
\mathbf{X}\|_2^2.\label{ju2}
\end{align}
Substituting \eqref{ju}-\eqref{ju2} into \eqref{Crr},
it then follows that
\begin{align}\label{CRB1}
&\!\!\!\!\mathrm{CRB}_{k}\left(\mathbf{q},\mathbf{u},  {\phi_k}\!\right)\nonumber\\
&\!\!\!\!=\frac{\sigma^2}{2|\rho_k|^2L}
    \left[\left\|\mathbf{h}_k(\mathbf{q},
    \mathbf{u},{\phi_k})
    ^H\mathbf{X}\right\|_2^2\left\|\dot{\mathbf{h}}_k(\mathbf{q},\mathbf{u},\phi_k)\right\|_2\right]^{-1}.
\end{align}

{\bf{Remark 1}}: From \eqref{CRB1}, we observe that the CRB for estimating the DoA of each target, $\mathrm{CRB}_{k}\left(\mathbf{q},\mathbf{u},  {\phi_k}\!\right)$, is critically dependent on the positions and rotations of the 6DMA surfaces ($\mathbf{q}$ and $\mathbf{u}$) through the corresponding LoS channels. Specifically, the potential CRB improvement originates from two main factors in \eqref{CRB1}, namely, the \emph{\textbf {power gain}}
 $\left\|\mathbf{h}(\mathbf{q}, \mathbf{u},{\phi}_k) ^H\mathbf{X}\right\|_2^2$ reflecting the probing power illuminating the target and the \emph{\textbf {geometric gain}} $\left\| \dot{\mathbf{h}}(\mathbf{q},\mathbf{u},{\phi}_k)\right\|_2$ reflecting the rate of change of the arc length (RAL) \cite{sal} of the antenna array, i.e., $\dot{\mathbf{h}}(\mathbf{q},\mathbf{u},{\phi}_k)$, which characterizes the impact of the 6DMAs' geometry on the target sensing accuracy.

\subsection{Problem Formulation}
Next, we aim at minimizing the CRB \footnote{Note that for the optimization of $\mathbf{q}$ and $\mathbf{u}$, we do not make the idealizing assumption that leads to \eqref{CRB1}.} in \eqref{CRB} for estimating the DoAs from all target locations by jointly optimizing the 3D positions $\mathbf{q}$ and 3D rotations $\mathbf{u}$ of all the 6DMA surfaces of the BS, subject to the practical movement constraints given in Section II-A. Accordingly, this optimization problem is formulated as follows:
\begin{subequations}
\label{MG3}
\begin{align}
\text{(P1)}~~&~\min\limits_{\mathbf{q},\mathbf{u}}
\mathrm{CRB}\left(\mathbf{q},\mathbf{u},  \boldsymbol{\phi}\right) \\
\text {s.t.}~&~\mathbf{q}_i\in\mathcal{C}, ~\forall i \in \mathcal{B}, \label{M1}\\
~&~ \|\mathbf{q}_i-
\mathbf{q}_{j}\|_2\geq d_{\min},~ 1\leq i < j \leq B,\label{M2}\\
~&~ \mathbf{n}(\mathbf{u}_i)^T(\mathbf{q}_{j}-\mathbf{q}_i)\leq  0,~ 1\leq i,j \leq B, i \neq j, \label{M3}\\
~&~\mathbf{n}(\mathbf{u}_i)^T\mathbf{q}_i\geq 0, ~\forall i\in \mathcal{B}.\label{jM3}
\end{align}
\end{subequations}
where constraint \eqref{M1} ensures that each 6DMA surface is positioned within the given 3D site space $\mathcal{C}$ of the BS.
Problem (P1) is non-convex due to the objective function's non-convexity w.r.t. the positions $\mathbf{q}$ and rotations $\mathbf{u}$ of the 6DMA surfaces, and the non-convex constraints in \eqref{M2}, \eqref{M3}, and \eqref{jM3}. The optimization problem is further challenged by the coupling between $\mathbf{q}$ and $\mathbf{u}$ in the objective function.

{\bf{Remark 2}}: We note that the optimal $\mathbf{q}$ and $\mathbf{u}$ in (P1) depend on the DoAs $\boldsymbol{\phi}$ of the typical targets in the different subregions. The locations of the actual targets, whose DoAs are to be estimated, will likely deviate from those of the typical targets. However, if the subregions are chosen sufficiently small, this deviation will be small and the optimized $\mathbf{q}$ and $\mathbf{u}$ will also be close-to-optimal for estimation of the DoAs of the actual targets.

\section{Proposed Algorithm}
\begin{algorithm}[t!]
\caption{PSO-Based Algorithm for Solving (P1).}
\label{alg3}
\begin{algorithmic}[1]
\STATE \textbf{Input}: $B$, $N$, $I$, $\{K_m\}_{m=1}^M$, and $T_{\mathrm{PSO}}$.  \\
\STATE \textbf{Output}: $\mathbf{q}$ and $\mathbf{u}$.  \\
\STATE Initialize particles with positions $\mathcal{R}^{(0)}$ and velocities
$\mathcal{V}^{(0)}$.
\STATE Obtain the local best position $\mathbf{s}_{\iota,pbest}=\mathbf{s}_{\iota}^{(0)}$ for $1 \leq \iota\leq I $ and the global best position\\
$\mathbf{s}_{gbest}=\arg \min\limits_{\mathbf{s}_{\iota}^{(0)}} \left\{ \mathcal{F}(\mathbf{s}_{1}^{(0)}),\mathcal{F}(\mathbf{s}_{2}^{(0)}),\cdots,
\mathcal{F}(\mathbf{s}_{I}^{(0)})\right\}$.
\FOR{$t = 1$ to $T_{\mathrm{PSO}}$}
\FOR{$\iota = 1$ to $I$}
\STATE Update the velocity of particle $\iota$ according
to \eqref{gb};
\STATE Update the position of particle $\iota$ according
to \eqref{vb};
\STATE Evaluate the fitness value of particle $\iota$, i.e.,
$\mathcal{F}(\mathbf{s}_{\iota}^{(t)})$
and update it according to \eqref{ada};
\IF {$\mathcal{F}(\mathbf{s}_{\iota}^{(t)})>\mathcal{F}(\mathbf{s}_{\iota,pbest})$}
\STATE Update $\mathbf{s}_{\iota,pbest}\leftarrow [(\mathbf{q}^{(t)})^T,(\mathbf{u}^{(t)})^T]^T$
\ENDIF
\IF {$\mathcal{F}(\mathbf{s}_{\iota}^{(t)})>\mathcal{F}(\mathbf{s}_{gbest})$}
\STATE Update $\mathbf{s}_{gbest}\leftarrow [(\mathbf{q}^{(t)})^T,(\mathbf{u}^{(t)})^T]^T$
\ENDIF
\ENDFOR
\ENDFOR
\STATE Obtain the suboptimal rotation and position $[\mathbf{q}^T,\mathbf{u}^T]^T=\mathbf{s}_{gbest}$.
\STATE Return $\mathbf{q}$ and $\mathbf{u}$.
\end{algorithmic}
\end{algorithm}

Inspired by the low complexity and high performance of the particle swarm optimization (PSO) algorithms
\cite{con}, we propose a PSO-based position and rotation optimization  scheme, which is described by Algorithm 1 for solving (P1). Algorithm 1 is an iterative procedure involving multiple particles, each characterized by two properties: position and velocity. The position of a particle, denoted by $\mathbf{s}=[\mathbf{q}^T, \mathbf{u}^T]^T\in\mathbb{R}^{6B\times 1}$, includes the parameters being optimized. The velocity of a particle, denoted by $\mathbf{v}\in\mathbb{R}^{6B\times 1}$, specifies how much the particle changes its position.

The objective of the optimization is to minimize the $\mathrm{CRB}\left(\mathbf{q},\mathbf{u},  \boldsymbol{\phi}\right)$, which is thus chosen as the fitness function; however, because of the constraints in (P1), the fitness function needs to be further adjusted \cite{con}. Specifically, in order to ensure constraints \eqref{M2}--\eqref{jM3} are met, we
introduce an adaptive penalty factor, which leads to the following fitness function:
\begin{align}\label{ada}
\mathcal{F}(\mathbf{s})=
\mathrm{CRB}\left(\mathbf{q},\mathbf{u},  \boldsymbol{\phi}\right)+
\tau|\mathcal{Q}(\mathbf{s})|_{\mathrm{c}},
\end{align}
where $\tau$ is a positive penalty parameter, and $\mathcal{Q}(\mathbf{s})$ denotes the penalty term for the infeasible particles. Specifically,
each entry of $\mathcal{Q}(\mathbf{s})$ represents a possible position-rotation pair of the 6DMA surfaces, i.e., a $(\mathbf{q},\mathbf{u})$ that violates the
minimum distance constraint \eqref{M2}, rotation constraint \eqref{M3}, or rotation constraint \eqref{jM3}. Hence, $\mathcal{Q}(\mathbf{s})$ can be defined as
\begin{align}\label{pr}
&\mathcal{Q}(\mathbf{s})=\{(\mathbf{q}_i,{\mathbf{q}}_j)|
\|{\mathbf{q}}_{i}-
{\mathbf{q}}_{j}\|_2^2< d_{\min}, 1 \leq i<j \leq B\} \nonumber\\
&\cup
\{(\mathbf{q}_j,\mathbf{u}_i)|\mathbf{n}
(\mathbf{u}_i)^T
(\mathbf{q}_{j}-\mathbf{q}_i)>  0,~ 1\leq i,j\leq B, i \neq j \} \nonumber\\
&\cup \{(\mathbf{q}_i,\mathbf{u}_i)|
\mathbf{n}(\mathbf{u}_i)^T\mathbf{q}_i< 0,~1\leq i\leq B\}.
\end{align}

The position and velocity of the $\iota$-th particle in iteration $t$ is denoted by  $\mathbf{s}_{\iota}^{(t)}$ and $\mathbf{v}_{\iota}^{(t)}$, respectively. As shown in Algorithm 1, in the initialization
process, we randomly initialize $I$ particles with the following positions and velocities, respectively:
\begin{align}
\mathcal{R}^{(0)}&=\{\mathbf{s}_1^{(0)},\mathbf{s}_2^{(0)},\cdots,
\mathbf{s}_I^{(0)}\},\label{rr0}\\
\mathcal{V}^{(0)}&=\{\mathbf{v}_1^{(0)},\mathbf{v}_2^{(0)},\cdots,
\mathbf{v}_I^{(0)}\}.\label{vv0}
\end{align}
Let $\mathbf{s}_{\iota,pbest}$ denote the
best position of the $\iota$-th particle, and $\mathbf{s}_{gbest}$ represent the global best position. Initially, we set $\mathbf{s}_{\iota,pbest}=\mathbf{s}_{\iota}^{(0)}$, $1 \leq \iota\leq I $, and $\mathbf{s}_{gbest}=\arg \min\limits_{\mathbf{s}_{\iota}^{(0)}} \left\{ \mathcal{F}(\mathbf{s}_{1}^{(0)}),\mathcal{F}(\mathbf{s}_{2}^{(0)}),\cdots,
\mathcal{F}(\mathbf{s}_{I}^{(0)})\right\}$.

Then, for each
iteration $t$, the velocity of each particle $\iota$ is updated as \cite{con}
\begin{align}\label{gb}
\mathbf{v}_\iota^{(t+1)}&=\kappa\mathbf{v}_\iota^{(t)}+
c_1\tau_1(\mathbf{s}_{\iota,pbest}
-\mathbf{s}_\iota^{(t)})+c_2\tau_2
\end{align}
where $c_1$ and $c_2$ are the individual and global learning factors, which determine the step size of each particle moving towards its best position. $\tau_1$ and $\tau_2$ are uniformly distributed random parameters in $[0, 1]$, which introduces randomness of the search to aid in escaping local minima. $\kappa$ represents the inertia weight, which maintains the particle's inertia during movement.
Subsequently, for each
iteration $t$, the position of each particle $\iota$ is updated as
\begin{align}
\mathbf{s}_\iota^{(t+1)}&=\mathcal{G}(\mathbf{s}_\iota^{(t)}+
\mathbf{v}_\iota^{(t)}),\label{vb}
\end{align}
with
\begin{align}\label{gg}
[\mathcal{G}({\mathbf{s}_\iota^{(t)}})]_j=\left\{\begin{matrix}
-\frac{A}{2}, ~~~\mathrm{if}~[\mathbf{s}_\iota^{(t)}]_j<-\frac{A}{2},\\
\frac{A}{2},~~~\mathrm{if}~[\mathbf{s}_\iota^{(t)}]_j>\frac{A}{2},\\
[\mathbf{s}_\iota^{(t)}]_j,~~~\mathrm{otherwise},
\end{matrix}\right.
\end{align}
for $1 \leq j \leq 3B$, where $A$ represents the side length of the BS region, which is assumed to be a cube in this letter. The projection function \(\mathcal{G}(\cdot)\) in \eqref{gg} ensures that the solution for problem (P1) always satisfies constraint \eqref{M1} during the iterations. Specifically, if a particle moves outside the boundary of the feasible region $\mathcal{C}$, its position is corrected by projecting the particle's component to the respective minimum or maximum value.

Next, we evaluate each particle's fitness based on \eqref{ada} and compare it against the fitness values at its local and global best positions. If a particle's fitness value surpasses that of its local best position or the global best position, the respective best positions are replaced with the current particle's position (refer to lines 8-14 in Algorithm 1).

The fitness value of the global best position is
non-increasing during the iterations in Algorithm 1, i.e., $\mathcal{F}(\mathbf{s}_{gbest}^{t+1})\leq \mathcal{F}(\mathbf{s}_{gbest}^{t})$.
Meanwhile, the objective value of problem (P1) is always
bounded. Thus, the convergence of the Algorithm 1 is
guaranteed. In addition, the complexity of PSO is determined by the number of particles and the required number of iterations for convergence \cite{con}, and is given by $\mathcal{O}(IT_{\mathrm{PSO}})$, where $T_{\mathrm{PSO}}$ represents the maximum number of iterations in Algorithm 1.

\section{Simulation}
In this section, we present simulation results to demonstrate
the effectiveness of the proposed design. We set $N=2$, $B=32$, and $\mathcal{C}$ is a cube with a side length of $A=0.6$ m.
We further consider three 2D circular sensing regions with $K_1=5$, $K_2=10$, and $K_3=15$, centered at distances of 20 m, 40 m, and 60 m from the BS center, with radii of $2$ m, $2\sqrt{2}$ m, $2\sqrt{3}$ m, respectively. The carrier frequency is 2.4 GHz. The antenna spacing on each 6DMA surface is $\frac{\lambda}{2}$ and $d_{\min}=\frac{\sqrt{2}}{2}\lambda+\frac{\lambda}{2}$. We set $[\mathbf{X}]_{:,l}$ as a circularly symmetric complex Gaussian random vector with zero mean and covariance $\mathbb{E}[[\mathbf{X}]_{:,l}[\mathbf{X}]_{:,l}^H]
=\frac{P}{NB}\mathbf{I}$, where $P$ denotes the transmit power. Note that the simulation results show the actual CRB in \eqref{CRB} instead of the simplified one in \eqref{CRB1}. In Algorithm 1, we set $I= 200$ and $T_{\mathrm{PSO}} = 300$.

For performance evaluation, we compare the proposed 6DMA with the following benchmark schemes, all utilizing a three-sector BS (a specific instance of the 6DMA with $B=3$ and approximately $\lceil\frac{NB}{3}\rceil$ antennas on each surface). Each sector antenna surface spans roughly $120^{\circ}$. 1) FPA: The 3D locations and 3D rotations of all antennas are fixed. 2) FA/MA \cite{9650760,zhulet,weidong,qingmove}: The rotations of all antennas remain unchanged, while we apply the proposed PSO-based algorithm to optimize the positions of all antennas within each sector antenna surface.
\begin{figure}[t!]
\vspace{-0.99cm}
\centering
\setlength{\abovecaptionskip}{0.cm}
\includegraphics[width=2.4in]{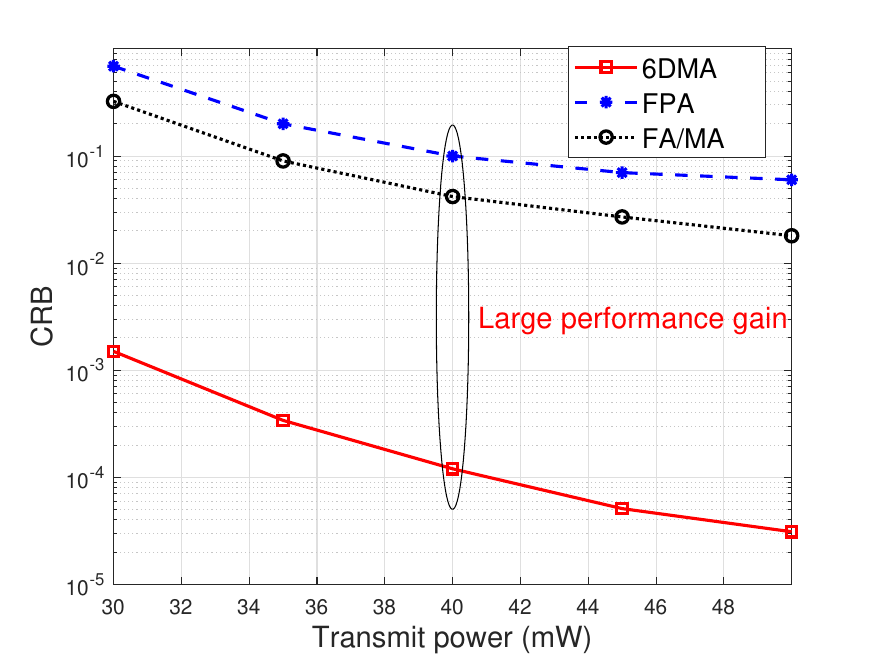}
\caption{CRB versus transmit power in scenario with directive antennas.}
\label{direc3}
\vspace{-0.69cm}
\end{figure}

Fig. \ref{direc3} shows the CRB versus the transmit power for a directive antenna pattern. According to the 3GPP standard, the  directive antenna gain $A(\tilde{\theta}_{b}, \tilde{\phi}_{b})$ in dBi is given by
\begin{align}\label{ji}
A(\tilde{\theta}_{b}, \tilde{\phi}_{b})=G_{\max}-\min\{-[A_{\mathrm{H}}
(\tilde{\phi}_{b})+A_{\mathrm{V}}(\tilde{\theta}_{b})],G_s\},
\end{align}
with $A_{\mathrm{H}}(\tilde{\phi}_{b})=-\min\{12\left(\frac{\tilde{\phi}_{b}}{\phi_{\mathrm{3dB}}}
\right)^2, G_s\}$ and $A_{\mathrm{V}}(\tilde{\theta}_{b})=-\min\{12(\frac{\tilde{\theta}_{b}}{\theta_{\mathrm{3dB}}}
)^2,G_v\}$. Here, $\theta_{\mathrm{3dB}}=\phi_{\mathrm{3dB}}=65^\circ$, $G_{\max}=8$ dBi, and $G_s=G_v=25$ dB. From Fig. \ref{direc3}, it is
observed that the 6DMA design performs much better than the FPA and FA/MA schemes, where the performance gaps become more
substantial as transmit power increases. This is due to the fact
that the 6DMA scheme has more spatial DoFs and can deploy the antenna resources more efficiently to match the subregions where targets are expected to be located. On the other hand, the FA/MA scheme can only adjust the antenna positions within each 2D surface. Therefore, the 6DMA scheme yields a higher power gain and a higher geometric gain for the targets' angles compared with the FPA and FA/MA schemes.

\begin{figure}[t!]
\vspace{-0.99cm}
\centering
\setlength{\abovecaptionskip}{0.cm}
\includegraphics[width=2.4in]{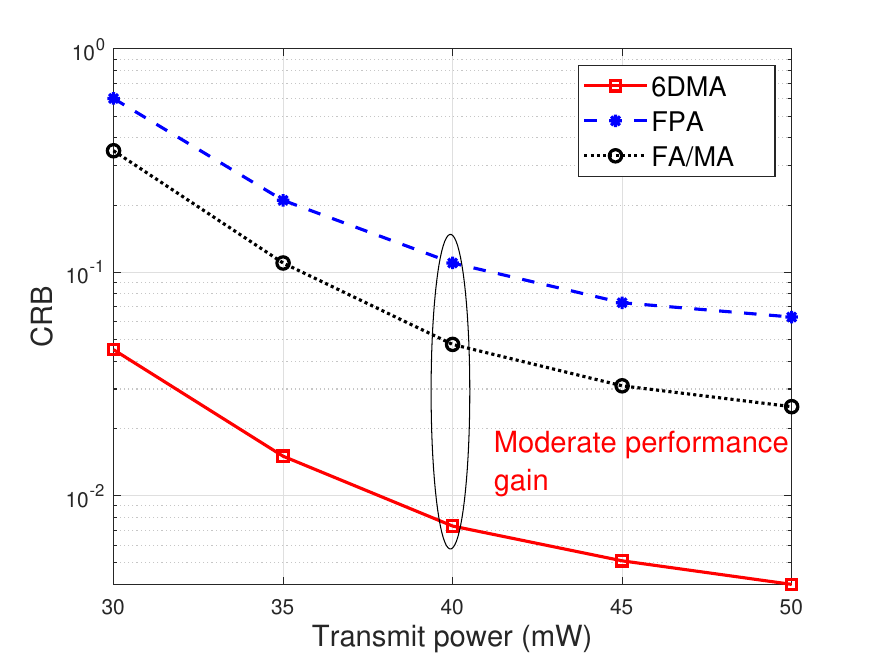}
\caption{CRB versus transmit power in scenario with isotropic antennas.}
\label{omni}
\vspace{-0.69cm}
\end{figure}

In Fig. \ref{omni}, we show the CRB versus the transmit power for a half-space isotropic antenna pattern, where
the antenna gain  in dBi is chosen as
\begin{align}\label{ho}
A(\tilde{\theta}_{b},\tilde{\phi}_{b})=\left\{\begin{matrix}
&10\log_{10}(2),&~ \text{if}~ -\frac{\pi}{2}\leq \tilde{\phi}_{b}\leq \frac{\pi}{2},\\
&-\infty,& ~ \text{otherwise.}
\end{matrix}\right.
\end{align}
From Fig. \ref{omni}, it is observed that the proposed 6DMA sensing scheme still outperforms the FPA and FA/MA schemes, although the performance gap is smaller compared to the case with directive antennas. This is because even with a half-space isotropic antenna pattern, the 6DMA scheme can achieve a moderate geometric/power gain by adjusting the positions and rotations of the 6DMAs to match
the transmitter-target LoS channels. For example, if a specific region contains a very large number of
subregions/targets, then the 6DMA system assigns more antennas for improved sensing in that area. In this way, the number of antennas are allocated to
different sensing regions, which leads to a balanced probing power illuminating the different targets, and thereby reducing the system's overall CRB as compared to the schemes without 6DMA.

\section{Conclusions}
This letter proposed a 6DMA-aided wireless sensing system and optimized its performance. To efficiently acquire the
typical target location information required for
6DMA position and rotation optimization, each sensing region was first divided into a number of equal-size subregions, and one typical target location within each subregion was selected.
Then, we studied a CRB minimization problem under practical movement constraints for 6DMAs. It was shown that CRB improvement of 6DMA sensing system can be attributed to a power gain and a geometric gain. Numerical results verified that for both directive and isotropic antenna radiation patterns, the proposed 6DMA design can significantly improve sensing accuracy compared to existing FPA and FA/MA approaches.

\bibliographystyle{IEEEtran}
\bibliography{fabs}
\end{document}